\documentclass[twocolumn,aps,prl,letterpaper]{revtex4}
\usepackage{graphicx}
\usepackage{amsmath}
\usepackage{amssymb}
\usepackage{wasysym}

\begin{document}  

\title{Charge Transport Scalings in Turbulent Electroconvection}
\author{Peichun Tsai,$^{1}$ Zahir A. Daya,$^{1,2}$ and Stephen W.
Morris$^{1}$}
\address{$^{1}$Department of Physics, University of
Toronto, 60 St. George St., Toronto, Ontario, Canada M5S 1A7\\
$^{2}$Defence Research \& Development Canada, 9 Grove Street, Dartmouth, Nova Scotia, Canada B2Y 3Z7}

\date{\today}

\begin{abstract}
We describe a local-power law scaling theory for the mean dimensionless 
electric current $Nu$ in turbulent electroconvection. The experimental system consists of a weakly conducting, submicron thick liquid crystal film supported in the annulus between concentric 
circular electrodes. It is driven into electroconvection by an applied voltage 
between its inner and outer edges. At sufficiently large voltage differences, 
the flow is unsteady and electric charge is turbulently transported between 
the electrodes. Our theoretical development, which closely parallels the 
Grossmann-Lohse model for turbulent thermal convection, predicts the 
local-power law $Nu \sim F(\Gamma) {\cal R}^{\gamma} {\cal P}^{\delta}$. ${\cal 
R}$ and ${\cal P}$ are dimensionless numbers that are similar to the Rayleigh 
and Prandtl numbers of thermal convection, respectively. The dimensionless function 
$F(\Gamma)$, which is specified by the model, describes the dependence of $Nu$ on the 
aspect ratio $\Gamma$. We find that measurements of $Nu$ are consistent with the theoretical model.
\end{abstract}

\maketitle

\section{Introduction}
Turbulent Rayleigh-B{\' e}nard convection (RBC), the paradigm for studies in 
convective turbulence, remains a fascinating unresolved puzzle~\cite{kadanoff_01}. For a century, 
theories of  turbulent RBC have focused on understanding the globally 
averaged heat transport through a layer of fluid.  The development of 
models has been largely driven by improved experimental measurements of 
the heat current, which have time and again revealed 
unexplained discrepancies between experiment and theory.  In response, mathematical models have become increasingly sophisticated during the last five decades~\cite{zaleski_98}. In recent years, ambitious experimental, theoretical and computational projects have been 
undertaken and the study of turbulent RBC has been 
considerably reinvigorated 
~\cite{ahlers_00,ahlers_side_00,niemela_00,GL_00,ahlers_01,GL_01,daya_ecke_01,
kerr_01,daya_ecke_02,xia_lam_zhou_02,verzicco_02,ahlers_03,niemela_03,GL_sidewall_03}.
All this activity suggests some clear directions for future work. 
Two recent experiments highlighted the crucial 
role of the system shape and lateral extent~\cite{daya_ecke_01,GL_sidewall_03}.  In this paper, we exploit the unique features of a system closely analogous to RBC, electrically driven convection in a thin annular film, to shed light on these features of turbulent convection.  

Turbulent RBC is described phenomenologically in terms of several 
 organizing structures which are experimentally observed. The convecting fluid has sharp thermal boundary layers at its top and bottom surfaces. Plumes grow erratically from these surfaces
and spontaneously organize into a noisy but coherent wind. This large
scale circulation (LSC), or turbulent wind, advects the plumes so that
hot thermals rise along a laterally bounding wall while cold plumes sink
along the diametrically opposite wall. The interior of the cell away
from the LSC is typically assumed to be well-mixed, isotropic and
homogeneous. This picture is partially corroborated by measurements using 
containers with approximately equal height and breadth. The extent to which this
phenomenological picture, particularly of the LSC, is 
geometry and shape dependent is currently the most pressing question in the study of turbulent RBC.  

The aspect ratio $\Gamma$, which is the ratio of the horizontal span to the 
vertical separation of an RBC apparatus, quantifies the geometry. It is important 
to perform experimental studies with a wide range of $\Gamma$. These have been difficult to accomplish however, due to practical considerations. The requirement of strong forcing,  which favors large vertical separation, is opposed by the constraint that the apparatus remain a manageable laboratory size, limiting its horizontal span. Consequently the 
majority of experiments have $\Gamma \approx 1$. Turbulent 
electroconvection, which is largely unencumbered by scale considerations, has emerged as a 
complementary experimental system for the study of convective turbulence 
~\cite{tsai_04}.  

\begin{figure}
\includegraphics[width=3in]{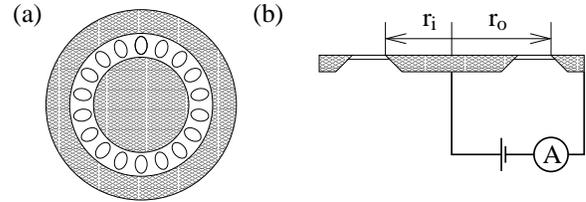}
\caption{\label{schematic}
Schematics of the experiment: (a) top and (b) side.}
\vspace{-0.2cm}
\end{figure}

\begin{figure}
\includegraphics[height=5.5cm]{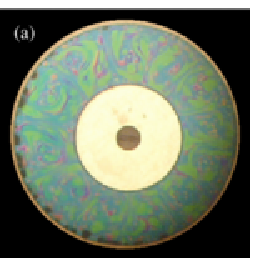}
\includegraphics[height=5.5cm]{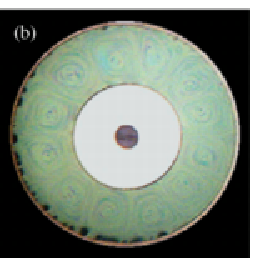}
\caption{\label{picture}
(Color online) Qualitative visualization of electroconvective flow, characterized by a film of uneven thickness: At (a) $V= 100$ volts and (b) $V = 250$ volts. Data, however, were acquired only from uniformly thick films in which no flow is visible.}
\vspace{-0.2cm}
\end{figure}

Electroconvection driven by DC potentials in thin smectic liquid crystal films has been extensively studied, mostly in the weakly driven, laminar regime~\cite{tsai_04, morris_90,mao_97,daya_97,dey_97,daya_98,langer_98,daya_99,daya_thesis_99,
daya_01,langer_01,daya_02}. The system consists of a smectic A 
liquid crystal film suspended in the annulus between the edges of concentric 
circular metallic electrodes as shown schematically in Fig.~\ref{schematic}. An 
applied electric potential difference between the inner and outer electrodes 
drives an electric current through the film. Surface charges accumulate on the 
two free surfaces that separate the electrically conducting film from 
charge-free space. This inverted surface charge density is unstable to 
electric forcing in much the same manner as the inverted mass density 
distribution of RBC is unstable to buoyancy forces. When the applied voltage $V$ 
exceeds the critical voltage $V_c$, the fluid is organized into convection
vortices as shown in Fig.~\ref{picture}a. The flow advects electric charge 
between the electrodes constituting a convection electric current. At higher 
driving, the flow becomes unsteady while retaining the large scale structure 
of convecting vortices, as shown in Fig.~\ref{picture}b. The turbulent electric 
charge transport is analogous to the heat flux in turbulent RBC.  

Other AC driven forms of electroconvection have been studied in bulk, three-dimensional liquid crystals~\cite{buka_kramer_95, ahlersEC, denninEC, kai_77, gleeson_01}.  In this paper, we concern ourselves only with two-dimensional, surface charge driven convection under DC potentials. However, it seems clear that the turbulent regime in these other systems might be amenable to a similar sort of scaling analysis~\cite{gleeson_01}.

The globally averaged heat transport in RBC is quantified in dimensionless form by dividing
out the contribution due to molecular conduction. It is then referred to as the
Nusselt number $Nu$. Almost every theoretical model over the last 5 decades has 
attempted as its central goal to describe the functional dependencies of $Nu$. It
is generally accepted that $Nu=Nu(Ra,Pr,\Gamma)$. Here $Ra$ is the Rayleigh number 
which quantifies the thermal forcing, $Pr$ is the Prandtl number which is the ratio of the 
fluid's kinematic viscosity to thermal diffusivity and $\Gamma$ is the aspect ratio
of the system.

Early models for $Nu$ used the
laterally extended limit, $\Gamma\rightarrow\infty$, while some recent
theories assume the confined case of $\Gamma \approx 1$~\cite{zaleski_98}.  A few years ago Grossmann and Lohse (GL) proposed a unifying theory for the $Ra$ and $Pr$ 
scaling of global variables like $Nu$~\cite{GL_00}, albeit for the case of $\Gamma \approx 1$.
The GL scheme begins by decomposing thermal turbulence into two constituents: a kinetic 
component and a thermal component. Then, using a second phenomenological distinction between the  boundary layer and the bulk, the GL theory estimates the kinetic and thermal 
dissipations in each region using dimensional arguments. Finally, GL theory 
derives the dependence of $Nu$ on $Ra$ and $Pr$ by balancing the exact formulations for the total 
dissipations versus the dominant contributions from the bulk and/or boundary layer. The various combinations of bulk/boundary 
layer, kinetic/thermal, and other considerations lead to the fragmentation of the $Ra-Pr$ 
parameter space into ten regimes~\cite{GL_01}. Within each regime, GL theory predicts the local-power law scalings 
$Nu\sim Ra^\gamma Pr^\delta$ with regime-dependent exponents $\gamma$ and $\delta$. Near regime boundaries, we expect crossover effects.  Unlike 
previous scaling theories, GL theory predicts no purely power law scaling for $Nu$. Rather, $Nu=f(Ra,Pr)$ can be approximated by regime-dependent combinations of 
local power laws for $\Gamma \approx 1$ systems. Recent precision experiments are better modeled by the GL cross-over scaling theory 
than by simple power laws~\cite{ahlers_00,ahlers_01,xia_lam_zhou_02}.        

Deep similarities between the phenomenologies and mathematical descriptions
of turbulent electroconvection and RBC make possible the development of a GL theory
for electroconvection. In this paper, we exploit these parallels to describe the
nondimensional electric current (also denoted $Nu$, the {\it electric} Nusselt number) in turbulent electroconvection.
We find that $Nu=F(\Gamma)f({\cal R},{\cal P})$ where ${\cal R}$ and ${\cal P}$ are the
electrical analogs of the Rayleigh and Prandtl numbers of conventional RBC.
The function $f({\cal R},{\cal P})$ can be approximated by local-power laws with
the same regime dependent exponents as in the GL theory. We find that the theoretical
predictions are consistent with electric current measurements. Unlike the GL model, where
it is assumed that $\Gamma \approx 1$, the naturally periodic annular geometry of electroconvection 
permits the explicit calculation of the aspect ratio dependence $F(\Gamma)$. The aspect ratio 
dependence of $Nu$ is not a power law as is the case in the theory of Shraiman and Siggia 
~\cite{shr_sig_90_94}, but instead approaches a constant for large $\Gamma$. We find surprisingly broad agreement between the function $F$ and measurements from both turbulent electroconvection
and RBC~\cite{tsai_04}. 

This paper is organized as follows. We start by describing the physical model
for annular electroconvection in thin freely suspended liquid crystal films. We 
then derive rigorous results for the globally averaged kinetic and electric dissipations.
Using two examples that are directly relevant to the current experiments, we show the 
derivation of local power law scaling for $Nu$ and for the Reynolds number $Re$ of the
large scale circulation. Important features of the
experimental system are then recounted and we describe direct comparison between the scaling theory  and experimental data. We finish with a brief conclusion where we discuss the complementary relationship between our work on turbulent electroconvection and conventional RBC.

\section{Scaling theory}
\label{theory}
\subsection{Background}

The basic equations for annular electroconvection are introduced in this section; for full details see Ref.~\cite{daya_99}.  A close parallel with RBC is evident, which allows us the develop a scaling theory, analogous to GL theory, in subsequent sections.

The annular film is a two dimensional (2D) sheet that in a cylindrical coordinate system 
spans the region $r_i \leq r \leq r_o$ at $z=0$. Here $r_i~(r_o)$ are the radii of the 
inner~(outer) electrodes. On the plane $z=0$, the inner~(outer) electrodes occupy the 
region $0 \leq r \leq r_i~(r_o \leq r \leq \infty)$. The film is a Newtonian, incompressible,
electrically conducting fluid with 2D density $\rho$, molecular viscosity $\eta$, and 
electric conductivity $\sigma$. We write the equations governing the fluid and charge 
in the annular film using 2D differential operators, field variables and material parameters.
The fluid flow is described by the Navier-Stokes equation with an electric body force:

\begin{equation}
\rho (\partial_t {\bf u}+ {\bf u} \cdot \nabla {\bf u} ) = -\nabla P + \eta \nabla^2 {\bf u} +
q {\bf E}\,.
\label{nseqn}
\end{equation} 
Here the fluid velocity ${\bf u} = (u,v)$, $P$ is the pressure, $q$ the surface charge density,
${\bf E}$ is the electric field and $q{\bf E}$ is the electric body force. The conservation
of charge leads to a continuity equation:

\begin{equation}
\partial_t q + \nabla \cdot {\bf J} = 0\,, \hspace{3mm} {\bf J} = \sigma {\bf E} + q {\bf u}\,.
\label{conteqn}
\end{equation}
The current density ${\bf J}$ is composed of the usual Ohmic or conduction density 
$\sigma {\bf E}$ and the convective current density $q {\bf u}$. The electric field and
charge density are not independent variables and have to satisfy Maxwell's equations.  Magnetic effects are negligible. 
In the region $|z| \neq 0$ {\em i.e.} above and below the film and electrodes, space is 
free of charges and so Laplace's equation for the electric potential holds:

\begin{equation}
{\nabla_3}^2 \psi_3 =0\,, \hspace{5mm} {\nabla_3}^2 = \nabla^2 + {\partial_z}^2\,.
\label{lapeqn}
\end{equation}
In the above, the subscript 3 identifies the potential and gradient operator are 
defined in three dimensions. The potential $\psi_3$ can determined in the upper half  space by solving 
Eqn.~\ref{lapeqn} subject to boundary conditions at $z=0$ and at infinity. The surface
charge density $q$ on the film due to the discontinuity in the electric field normal to
the film is given by

\begin{eqnarray}
\nonumber q&=&-\epsilon_0 \partial_z \psi_3 |_{z=0^+} + 
\epsilon_0 \partial_z \psi_3 |_{z=0^-}\\
&=& -2\epsilon_0 \partial_z \psi_3 |_{z=0^+}\,, \hspace{5mm} r_i \leq r \leq r_o\,.
\label{qeqn}
\end{eqnarray}
In the above $\epsilon_0$ is the permittivity of free space. Equations \ref{nseqn}-\ref{qeqn} 
model the electroconvection system. The equations
are subject to rigid boundary conditions on the fluid velocity: ${\bf u}=0$ at $r=r_i$ 
and $r=r_o$ and to Dirichlet boundary conditions for the electric potential at $z=0$ and
at infinity. The applied potential is $V$~volts for $r \leq r_i$ and $0$ volts for 
$r \geq r_o$ and at infinity. On the film, the electric potential is determined by
satisfying the current density ${\bf J}$ and the boundary conditions at $r=r_i,~r_o$. Using
the Dirichlet Green function~$G(r,\theta,z;r',\theta',z')$, we can formally solve Eqns.
~\ref{lapeqn} and~\ref{qeqn} for the surface charge density:

\begin{eqnarray}
q(r,\theta) &=& \frac{\epsilon_0}{2\pi}\biggl[\oint da'~\psi_3 (r',\theta',z'=0)
\frac{\partial^2 G}{\partial z \partial z'}\bigg|_{z'=0}\biggr]_{z=0^+} \nonumber\,,\\
& \equiv & g[\psi]\,, \hspace{5mm} \psi = \psi_3(r,\theta,z=0)\,.
\label{defg}
\end{eqnarray}
The integral is over the bounding surface, in this case the $z=0$ plane and surface at 
infinity. Eqn.~\ref{defg} essentially defines $q$ in terms of a functional $g$ whose
argument is the electric potential in the plane $z=0$. On this plane the electric field ${\bf E}$ and 
potential $\psi$ are related though ${\bf E}= -\nabla \psi$. Using this relation,
the definition of $g$, denoting the kinematic pressure field $p=P/\rho$ and kinematic 
viscosity $\nu=\eta/\rho$, the four Eqns.~\ref{nseqn}-\ref{qeqn} reduce to the 
following pair:

\begin{eqnarray}
{\partial}_t {\bf u} + {\bf u} \cdot \nabla {\bf u} &=& -\nabla p +
\nu \nabla^2 {\bf u} - \frac{g[\psi]}{\rho} \nabla \psi \,, \label{ns} \\
{\partial}_t \psi + {\bf u} \cdot \nabla \psi &=& \frac{\sigma {\nabla}^2 \psi}{\partial_{\psi}g }
\label{ct}\, .
\end{eqnarray}

Written in this way, the Eqns.~\ref{ns} and \ref{ct} bear striking similarity to
the Boussinesq equations for turbulent RBC with the scalar temperature and electric 
potential fields assuming similar roles. The solutions of Eqns.~\ref{ns} and \ref{ct}  are subject to the usual no-slip boundary conditions on 
${\bf u}$ and the applied electric potential boundary conditions on $\psi$. Three dimensionless parameters describe
the state of the system. ${\cal R}$, the analog of the Rayleigh number in RBC, is the control or external driving parameter,  and is proportional to the square of the applied
voltage.  It is given by 
\begin{equation}
{\cal R}=\frac{\epsilon_0^2 V^2}{\sigma \eta}\,.
\label{Rdef}
\end{equation} 
${\cal P}$, the analog of the Prandtl number in RBC, is the ratio of the charge to viscous 
relaxation time scales in the film:
\begin{equation}
{\cal P}=\frac{\epsilon_0 \eta}{\rho \sigma (r_o-r_i)}\,.
\label{Pdef}
\end{equation}
The geometry is uniquely characterized by the radius ratio $\alpha = r_i/r_o$. However, in order to make comparison to RBC, 
it is more appropriate to describe the geometry in terms of the aspect ratio $\Gamma$, 
which is the ratio of the horizontal or lateral dimension to the vertical
or transverse dimension. Since the lateral dimension is ambiguous in the annular geometry, $\Gamma$ is not uniquely determined.  However, $\Gamma$ can be consistently taken to be the ratio 
of the circumference of the film measured at the inner electrode to the film width, so that 
\begin{equation}
\Gamma=\frac{2\pi r_i}{r_o-r_i}=\frac{2\pi \alpha}{1-\alpha}\,.
\label{ARdef}
\end{equation}
Two possible alternative definitions of the aspect ratio use the mid-radius 
circumference   or the outer electrode circumference  as the lateral dimension, leading to $\Gamma_m$ and $\Gamma_o$, respectively. The  
two aspect ratios so defined are related to $\Gamma$
defined in Eqn.~\ref{ARdef} as $\Gamma=\Gamma_m - \pi = \Gamma_o - 2\pi$. Since the
various definitions of the aspect ratio are very similar, we have chosen to use the
form in Eqn.~\ref{ARdef} because in that case $0 \leq \Gamma \leq \infty$, as is true  for conventional
RBC systems. The alternative definitions $\Gamma_m$ and $\Gamma_o$ have $\pi$ and $2\pi$ as their lower bounds.  In any case, direct comparisons between rectangular and annular systems converge in the $\alpha \rightarrow 1$ or $\Gamma \rightarrow \infty$ limits.

\subsection{Global averages}

In this section we derive exact expressions for various globally averaged quantities, which, according to the GL procedure, will be balanced against scaling estimates, as discussed in the following section.

The convective contribution to the electric current is determined by
dividing out the conduction current from the total current. This dimensionless ratio 
is the Nusselt number $Nu$. The net electric current is radial between the inner 
and outer electrodes with a net zero contribution from azimuthal currents. Thus at 
any radial position $r_i \leq r \leq r_o$, the integral over the azimuth of the
radial component of the current density is equal to the total electric current. Hence, we 
find 

\begin{equation}
Nu \equiv \frac{\oint {\bf J}\cdot {\bf \hat r}~dl}{\oint {\bf J_{cond}}\cdot {\bf \hat r}~dl}=
\frac{\oint uq-\sigma \partial_r\psi~dl}{\oint -\sigma \partial_r\psi_{cond}~dl }.
\label{Nudef}
\end{equation}  
In the above, we have used the definition of ${\bf J}$ given in Eqn.~\ref{conteqn}.
The contour $dl$ is a circle at radius $r$, and $\bf \hat r$ is a radially outward unit vector. Note that $u$ is the radial component 
of the fluid velocity ${\bf u}$ and 
$\psi_{cond}$ is the electric potential for the conductive state. On the 
inner electrode ($r \leq r_i$) $\psi_{cond}=V$ while on the outer electrode 
($r \geq r_o$) $\psi_{cond}=0$. On the film where $r_i \leq r \leq r_o$, 
$\psi_{cond}$ is given by
\begin{equation}
\psi_{cond}=\frac{V}{\ln{\alpha}}\ln \frac{r}{r_o}\,.
\label{psicond}
\end{equation} 
Evaluating the terms in Eqn.~\ref{Nudef} using Eqn.~\ref{psicond}, we find the
relation
\begin{equation}
Nu = \frac{\int_0^{2\pi}(uq-\sigma {\partial}_r\psi)~rd\theta}{2 \pi \sigma 
V/\ln(1/\alpha)}\,.
\label{Nufinal}
\end{equation}

We use the angular brackets $\langle \cdot \cdot \cdot  \rangle$ to denote averages
over the fluid volume (actually its area), we write
\begin{equation}
\langle \cdots 
\rangle=\frac{\int~\cdots~rdrd\theta}{\int~rdrd\theta}=
\frac{\int_0^{2\pi} \int_{r_i}^{r_o}~\cdots~rdrd\theta}{\pi (r_o^2 - r_i^2)}\,.
\end{equation} 
As in the GL theory for RBC, our charge transport scaling theory begins with the 
kinetic and electric dissipation rates
\begin{eqnarray}
\epsilon_{\bf u}(r,\theta,t) &=& \nu (\nabla {\bf  u})^2 \nonumber \\
\epsilon_\psi (r,\theta,t) &=&  \sigma (\nabla \psi)^2 \,.
\end{eqnarray}
We denote the averages of the above dissipations over the fluid volume as
\begin{eqnarray}
\epsilon_{\bf u} &=& \langle \epsilon_{\bf u}(r,\theta,t) \rangle 
=\langle \nu (\nabla {\bf  u})^2 \rangle \nonumber \\
\epsilon_\psi &=&\langle \epsilon_\psi (r,\theta,t) \rangle 
=\langle \sigma (\nabla \psi)^2 \rangle \,.
\label{dissip}
\end{eqnarray} 
The rather complex and nonlocal functional relationship $g[\psi]$ between $q$ and $\psi$ given in 
Eqn.~\ref{defg} makes it very difficult to calculate the globally averaged
kinetic and electric dissipations. However, by splitting the integrand in Eqn.~\ref{defg} into local and nonlocal parts, $g$ can be expanded as
\begin{equation}
g[\psi] = \frac{\epsilon_0}{r_o-r_i}\psi + \tt{nonlocal~terms}\,. 
\label{gexpan}
\end{equation}
To a first approximation the functional $g[\psi]$ 
can be approximated as a linear function $g(\psi)$ so that the surface charge density and the
electric potential on the film are related {\em locally}. Then 
$\partial_{\psi} g$ is constant and Eqn.~\ref{ct} is identical to the
heat equation in RBC, with electric potential in place of temperature. The critical 
parameters and the critical mode numbers at the onset of electroconvection have been successfully 
captured by this local approximation, which is described in detail in 
Ref.~\cite{daya_99}. We assume that  $g \propto \psi$ continues to hold  at much higher forcing and develop the GL theory for turbulent electroconvection within the local approximation, which makes Eqns.~\ref{ns} and \ref{ct} very similar to the Boussinesq equations for thermal convection. The electric potential in electroconvection
almost precisely corresponds to the temperature in RBC.

The velocity and electric potential fields share the azimuthal periodicity of 
the annulus and thus permit the calculation of the globally-averaged kinetic and 
electric dissipations of Eqn.~\ref{dissip}. Assuming time-stationarity for the 
spatial averaging, we find from Eqns.~\ref{ns} and \ref{ct} the following relations for the kinetic dissipation ${\epsilon}_{\bf u} \equiv 
\langle \nu (\nabla {\bf  u})^2 \rangle $ and  electric dissipation 
${\epsilon}_\psi \equiv\langle \sigma (\nabla \psi)^2 \rangle$:

\begin{eqnarray}
{\epsilon}_{\bf u} &=& \frac{\nu^3 {\cal R} {\cal 
P}^{-2} ({\rm Nu} - 1)}{\ln(1/\alpha)(r_o^2-r_i^2)(r_o-r_i)^2} \label{eu}\,,\\
{\epsilon}_\psi &=& \frac{2\sigma V^2 {\rm Nu}}{\ln(1/\alpha)(r_o^2-r_i^2)} \label{ep} \,.
\end{eqnarray}

The above relations have interesting similarities and differences to those for the
corresponding quantities in RBC, which are given in Refs.~\cite{GL_00,GL_01,shr_sig_90_94}. The differences are entirely due to 
the annular geometry of the electroconvection system. In the narrow gap limit, in which the 
radius ratio $\alpha \rightarrow 1$ while film width $d=r_o-r_i$ remains constant, 
Eqns.~\ref{Nufinal},~\ref{eu} and \ref{ep} recover the familiar forms for RBC between
parallel plates.

\subsection{Grossmann-Lohse scalings} 

In this section, we make several assumptions about the spatial organization of the turbulent flow, in the same manner as in the scaling theory of Grossmann and Lohse (GL), as they explain in Section 2 of Ref.~\cite{GL_00}. In particular,
GL assumed that a turbulent wind, or LSC, comprising a single cell occupies the 
entire $\Gamma \sim 1$ RBC container. The wind is driven by plumes from the boundary layers and it in turn 
drives the interior or bulk. With the boundary layers and bulk conceptually distinguished,
GL estimate the relative boundary and bulk dissipations. Here, we make analogous  
assumptions about turbulent electroconvection. As shown in Fig.~\ref{picture}b, the
turbulent flow consists of counter-rotating convection vortices around the annulus.
The vortices, which are unsteady, have fluctuating boundaries that are defined by 
the averaged turbulent LSC. Each vortex is assumed to be roughly square with  
dimension $r_o-r_i$. Near the electrodes, well developed viscous and electric 
boundary layers with respective thicknesses $\lambda_{\bf u}$ and $\lambda_\psi$ 
are assumed. Away from the electrodes, the vortex  interior, or bulk, is taken to be 
well mixed. We do not account for the slight differences between the boundary 
layer dimensions at the inner and outer electrodes due to the annular geometry.
This asymmetry diminishes with increasing aspect ratio and is assumed to be always
small. 

The total dissipations calculated in Eqns.~\ref{eu} and \ref{ep} are decomposed 
into contributions from the boundary and bulk regions of the convection cells as follows

\begin{eqnarray}
\epsilon_{\bf u} &=& \epsilon_{\bf u}^{bl} + \epsilon_{\bf u}^{bulk} \,,\nonumber \\
\epsilon_\psi &=& \epsilon_\psi^{bl} + \epsilon_\psi^{bulk} \,.\label{dissdecom}
\end{eqnarray}
The contribution $\epsilon_{\bf u}^{bl}$ of the boundary layer kinetic dissipation
is defined as
\begin{eqnarray}
\epsilon_{\bf u}^{bl} &=& \langle \epsilon_{\bf u}(r,\theta,t) \rangle_{bl} 
=\langle \nu (\nabla {\bf  u})^2 \rangle_{bl}\,, \nonumber \\
\langle \cdots \rangle_{bl}&=& \frac{\int_0^{2\pi} \int_{r_i}^{r_i + \lambda_{\bf u}}~\cdots~rdrd\theta}
{\pi((r_i + \lambda_{\bf u})^2 - r_i^2)}\nonumber \\ 
&+& \frac{\int_0^{2\pi} \int_{r_o-\lambda_{\bf u}}^{r_o}~\cdots~rdrd\theta}
{\pi(r_o^2 - (r_o - \lambda_{\bf u})^2)} \,.\label{blave}
\end{eqnarray}
The electric dissipation in the bulk $\epsilon_\psi^{bulk}$ is defined as
\begin{eqnarray}
\epsilon_\psi^{bulk} &=& \langle \epsilon_\psi(r,\theta,t) \rangle_{bulk} 
=\langle \sigma (\nabla \psi)^2 \rangle_{bulk}\,, \nonumber \\
\langle \cdots \rangle_{bulk}&=& \frac{\int_0^{2\pi} \int_{r_i+\lambda_\psi}^{r_o - \lambda_\psi}
~\cdots~rdrd\theta}{\pi((r_o-\lambda_\psi)^2 -(r_i + \lambda_\psi)^2)} \,. \label{bulkave}
\end{eqnarray} 
The other two boundary and bulk dissipations are similarly defined. We assume that 
\begin{equation}{\rm Nu}={\rm F}(\Gamma) {\rm f}({\cal R},{\cal P})\,.\label{NuFf}
\end{equation}
It is our purpose to determine the as yet unspecified functions ${\rm F}$ and ${\rm f}$. 
We begin by estimating the bulk and boundary layer dissipations. In the following we 
ignore numerical factors of $O(1)$.

\subsubsection{Kinetic dissipations}
The turbulent wind or LSC sets the velocity scale $U$ for both the boundary and bulk regions. 
A viscous or kinetic boundary layer, assumed to be laminar, scales as $\lambda_{\bf u} 
\sim (r_o-r_i) Re^{-1/2}$. Here $Re$ is the Reynolds number based on velocity $U$ and 
film width $r_o-r_i$.
Finally we assume that $\lambda_{\bf u} \ll r_i$, $\lambda_{\bf u} \ll r_o$ and
$\lambda_{\bf u} \ll r_o-r_i$ which greatly simplifies the averaging in Eqns.~\ref{blave} and 
\ref{bulkave}. For the boundary layer, we estimate the kinetic dissipation by
\begin{eqnarray}
\epsilon_{\bf u}^{bl} &=& \langle \nu (\nabla {\bf  u})^2 \rangle_{bl} \sim 
\nu \biggl(\frac{U}{\lambda_{\bf u}}\biggr)^2\frac{\lambda_{\bf u}}{r_o-r_i} \,,\nonumber \\
&\sim& \frac{\nu^3}{(r_o-r_i)^4}Re^{5/2}\,. \label{eubl}
\end{eqnarray}
The LSC stirs the interior and so, following GL,  we use the convective term to estimate the bulk
kinetic dissipation  
\begin{eqnarray}
\epsilon_{\bf u}^{bulk} &=& \langle \nu (\nabla {\bf  u})^2 \rangle_{bulk} \sim 
{\bf u}\cdot ({\bf u}\cdot \nabla){\bf u}\sim \frac{U^3}{r_o-r_i}\,,\nonumber \\
&\sim& \frac{\nu^3}{(r_o-r_i)^4}Re^3\,. \label{eubulk}
\end{eqnarray}
The relations \ref{eubl} and \ref{eubulk} are virtually identical to their counterparts
for turbulent RBC given in Section 2.3 of Ref.~\cite{GL_00}.

\subsubsection{Electric dissipations}
The applied potential $V$ drops over the electric potential boundary length $\lambda_\psi$,
which we assume is small such that $\lambda_\psi \ll r_i$, $\lambda_\psi \ll r_o$ and
$\lambda_\psi \ll r_o-r_i$. As with the kinetic dissipations, these requirements greatly
simplify the averaging prescribed by Eqns.~\ref{blave} and \ref{bulkave}. Since most of the 
applied potential drops over $\lambda_\psi$ at the inner and outer electrodes, there is
effectively an electrical short {\em i.e.} a constant potential, in the bulk. Invoking the standard
arguments of RBC, where the thermal short determines a relation between the thermal boundary layer
and the heat transport, we find that 
\begin{equation}
\lambda_\psi \sim \frac{(r_i+r_o) \ln{(1/\alpha)}}{Nu} \label{elecshort} \,.
\end{equation}
Here, we have assumed that half the applied potential drops over the potential boundary layer
at the inner electrode and half at the outer electrode. We assume that the electric potential
boundary layer thicknesses are the same at the inner and outer electrodes. This symmetry is 
exact for radius ratio $\alpha \rightarrow 1$ or aspect ratio $\Gamma \rightarrow \infty$. Our
derivation implicitly assumes that the role of the asymmetry is not crucial at smaller $\Gamma$.

In the boundary layer, the electric dissipation is given by
\begin{equation}
\epsilon_\psi^{bl} = \langle \sigma (\nabla \psi)^2 \rangle_{bl} \sim 
\sigma \biggl(\frac{V}{\lambda_\psi}\biggr)^2\frac{\lambda_\psi}{r_o-r_i} \,.\label{epsibl}
\end{equation}
By balancing the latter two terms in Eqn.~\ref{ct}
\begin{equation}
{\bf u} \cdot \nabla \psi \approx \frac{\sigma {\nabla}^2 \psi}{\partial_{\psi}g }\,,
\end{equation}
we find, up to a linear expansion of $g(\psi)$, that
\begin{equation}
\frac{V}{\lambda_\psi} \sim \frac{\sigma (r_o-r_i) V}{\epsilon_0 U \lambda_\psi^2}\,.\label{sup}
\end{equation}
By using $U$ as the velocity scale, we have implicitly assumed that $\lambda_\psi \leq \lambda_{\bf u}$.
Substituting in Eqn.~\ref{epsibl} for $V/\lambda_\psi$ from Eqn.~\ref{sup} and for $\lambda_\psi$ from
Eqn.~\ref{elecshort} we get
\begin{equation}
\epsilon_\psi^{bl} \sim \biggl(\frac{1+\alpha}{1-\alpha}\biggr)\frac{\ln(1/\alpha)~Re~{\cal P}\sigma V^2}
{Nu~(r_o-r_i)^2} \,.\label{epsiblF}
\end{equation}

The dissipations estimated in Eqns.~\ref{eubl},~\ref{eubulk} and \ref{epsiblF} are sufficient to define the two relevant regimes for turbulent electroconvection. In one scenario,
we assume that both the electrical and kinetic dissipations occur primarily in the boundary. Then
Eqn.~\ref{dissdecom} will be written as
\begin{eqnarray}
\epsilon_{\bf u} &\approx& \epsilon_{\bf u}^{bl} \,,\nonumber \\
\epsilon_\psi &\approx& \epsilon_\psi^{bl}\,.\label{blbl}
\end{eqnarray}
This corresponds to the ${\rm I}_l$ regime of Ref.~\cite{GL_01}. For the left-hand-sides in the above equations, we use the globally averaged dissipations derived 
from the equations of motion given in Eqns.~\ref{eu} and \ref{ep}. For the right-hand-sides we use the
boundary layer estimates given in Eqns~\ref{eubl} and \ref{eubulk}. After some algebraic manipulation, we
find 
\begin{equation}
Nu \propto F(\Gamma) {\cal R}^{1/4} {\cal P}^{1/8}~~~({\rm regime~I}_l~{\rm of~Ref.~[8]}).
\end{equation}
Repeating the above procedure for   
\begin{eqnarray}
\epsilon_{\bf u} &\approx& \epsilon_{\bf u}^{bl} \,,\nonumber \\
\epsilon_\psi &\approx& \epsilon_\psi^{bulk}\,,\label{blbulk}
\end{eqnarray}
which corresponds to regime ${\rm II}_l$ of Ref.~\cite{GL_01}, we find that 
\begin{equation}
Nu \propto F(\Gamma) {\cal R}^{1/5} {\cal P}^{1/5}~~~({\rm regime~II}_l~{\rm of~Ref.~[8]}).
\end{equation}

 In general, this procedure gives a power law dependence for $\rm f$ in 
Eqn.~\ref{NuFf} of the form 
\begin{equation}
{\rm f} \sim {\cal R}^{\gamma} {\cal P}^{\delta}\,.\label{fpower}
\end{equation}
It also follows that the Reynolds number ${\rm Re}$ of the vortices, based on the radial velocity of the fluid and the separation between the electrodes $r_o-r_i$,  scales as
\begin{equation}
{\rm Re} \sim {\cal R}^{\gamma_\ast}{\cal P}^{\delta_\ast}\,.\label{Repower}
\end{equation}
We find the same set of exponents $\gamma,\delta,\gamma_\ast,\delta_\ast$ as the GL theory for turbulent RBC.  The ${\rm I}_l$ and ${\rm II}_l$ regimes are relevant to the low $\cal R$ and broad $\cal P$ range that can be accessed by smectic electroconvection.

\subsubsection{aspect ratio dependence ${\rm F}(\Gamma)$}

After balancing the dominant contributions, as in Eqns.~\ref{blbl} and \ref {blbulk}, we find that the power law 
${\cal R}^\gamma {\cal P}^\delta$ has a common algebraic prefactor ${\rm F}$ that is only a function of 
the aspect ratio $\Gamma$. Unlike in previous studies of RBC in Cartesian geometry~\cite{shr_sig_90_94}, ${\rm F}(\Gamma)$ is not 
itself a power law.   Instead, it is given by
\begin{equation}
{\rm F}(\Gamma)= \frac{\Gamma + \pi}{\pi} \ln\Biggl(\frac{\Gamma+2\pi}{\Gamma}\Biggr)\,.\label{Fdef}
\end{equation}
${\rm F}(\Gamma)$ specifies the aspect ratio dependence of the global charge or heat transport, as contained in the Nusselt number $Nu$.  The Reynolds number ${\rm Re}$ of the  large scale circulation, however, is local to each vortex and is independent of $\Gamma$.

\section{Experiment and results}

In this section, we describe the main features of the experimental apparatus and procedure, the data analysis and the results.  Our apparatus is similar to one used previously for studies of thin film electroconvection in the weakly nonlinear regime~\cite{morris_90,mao_97,daya_97,dey_97,daya_98,langer_98,daya_99,daya_thesis_99,
daya_01,langer_01,daya_02}, which has more recently been adapted to the turbulent regime~\cite{tsai_04}. Other forms of electroconvection in bulk liquid crystals have similarly been extended to study scaling in the turbulent regime~\cite{kai_77, gleeson_01}.

The experiment consists of an annular liquid crystal film of octylcyanobiphenyl~(8CB) freely suspended between two concentric stainless steel electrodes as shown in Fig.~\ref{schematic}. The annular film was about $2~cm$ in diameter. The experiment is enclosed by an aluminum box which serves both as a faraday cage and as a rough vacuum chamber. For the experiments discussed here, the film was at atmospheric pressure and was temperature controlled to $24\pm 2^\circ$C. At this temperature, 8CB is in the smectic A phase. In this phase, the elongated  molecules align normal to the plane of the film, which consists of an integer number of layers.  In 8CB, each layer is $3.16~nm$ thick~\cite{thickness}. In all experiments, we used films that were uniformly thick to within $\pm 3$ layers, and which had thicknesses between $30$ and $85$ layers.  Within the plane of the layers, the film closely approximates an ideal 2D Newtonian fluid. The layered structure strongly restricts fluid motion perpendicular to the layers. The film thickness is comparable to the wavelength of visible light. We determined the thickness of a film by its interference color under reflected white light, using standard colorimetric functions~\cite{handbook_optics, colour_dynamics}. During the experiments we observed and recorded the reflected film color with a CCD camera. 

Pure 8CB has a low, uncontrolled electrical conductivity due to residual ionic impurities. To control the conductivity, we dope the 8CB with tetracyanoquinodimethane~(TCNQ)~\cite{tcnq}, an electron acceptor. With a concentration of TCNQ between $5\times 10^{-5}$ and $5\times 10^{-4}$ by mass, we find that the sample has a bulk conductivity in the range $10^{-8}$ to $10^{-7}~ \Omega^{-1}{\rm m}^{-1}$. We determine the conductivity of each film from its ohmic response below the onset of convection, as discussed below.

The experimental procedure consists of applying a DC voltage $V$ across the film and measuring the resulting electrical current $I$. The inner (outer) electrode is electrically high (grounded). The applied voltage is incremented in small steps from $0$ to $1000$~V and then decremented to zero. At each voltage, the current is measured with a computer-interfaced electrometer, which is equipped with low noise triaxial cables. The film resistance is in the ${\rm T}\Omega$ range and typical currents are $\sim 1~{\rm pA}$. To determine the dimensionless charge transport $Nu$, we require an accurate value for the critical voltage $V_c$ at the onset of convection. To measure $V_c$, we use a small voltage step, $\sim 1$ volt in the voltage range between $0$ and $50$~V, which brackets the typical critical voltage for most of our films. A larger voltage step of $\sim 10$~V is used in the range $60$ to $1000$~V. The larger step is necessary to limit the drift in electrical conductivity due to electrochemical reactions in the film. At each applied voltage, we make $100$ current measurements spaced by $25$~ms. The average values of the current were used to calculate $Nu$, as described below.  In addition to the $IV$ data, we measured the film thickness $s$ and aspect ratio $\Gamma$. 
Further details of the material preparation and experimental procedure can be found in Refs.~\cite{dey_97, daya_99, daya_01}.

From the slope of the $IV$ curve in the conduction regime where the film is quiescent, we determined the film conductance $C$. 
The Nusselt number is the current divided by the conduction current, $Nu~=~I/I_c=~I/CV$. The control parameter ${\cal R}$ defined in Eqn.~\ref{Rdef}, was calculated using the accepted values of the viscosity, the measured conductivity, film thickness and the applied voltage. It varies
between $0$ and $\sim 10^5$. This is moderate compared with the very high Rayleigh numbers achievable in RBC. However, the critical value ${\cal R}_c$ of ${\cal R}$ at the onset of electroconvection is about a factor of $10$ smaller than that for RBC. The Prandtl number ${\cal P}$ defined in Eqn.~\ref{Pdef} was calculated from the material parameters and film dimensions. In the experiments reported here,  $5 < {\cal P} < 250$.

Electroconvecting annular smectic films have several advantages for the study of turbulent scaling, relative to conventional RBC. In RBC experiments, the heat losses through the sidewalls must be taken into account and corrected for~\cite{ahlers_side_00}, whereas the annular film has no sidewalls.  The characteristic time scales of electroconvection are many orders of magnitude shorter than in RBC, making data acquisition much faster.  Also, the annular aspect ratio can be easily varied over a broad range ($0.3~\leq \Gamma~\leq 17$ for the experiments reported here). A similar range of $\Gamma$ in RBC is possible in principle, but  would be very cumbersome and time consuming in practice. 

Electroconvecting smectic films also have some clear disadvantages, relative to RBC.   Degradation of the liquid crystal under DC excitation results in significant conductivity drifts over an experimental run. The drift can be as large as $30\%$ over the course of an experiment.  The drift can be partially compensated for by monitoring the change in the critical voltage $V_c$ before and after each sweep of the voltage, as discussed further below.  The total DC charge flow is quite large; $\sim 0.4\mu$C over a time interval of $\sim 1$h. The conductivity drift results in an uncontrolled variation of both ${\cal R}$ and ${\cal P}$, increasing the uncertainty in these parameters.  A second disadvantage is the rather modest upper limit on ${\cal R}$, relative to that attainable in RBC.  Higher ${\cal R}$ could be had by simply increasing the applied voltage $V$. However, large electric fields will eventually lead to dielectric breakdown, destroying the film. We estimate that with the present material parameters, the maximum accessible ${\cal R}$ is about $10^6$.

\begin{figure}
\includegraphics[height=6cm]{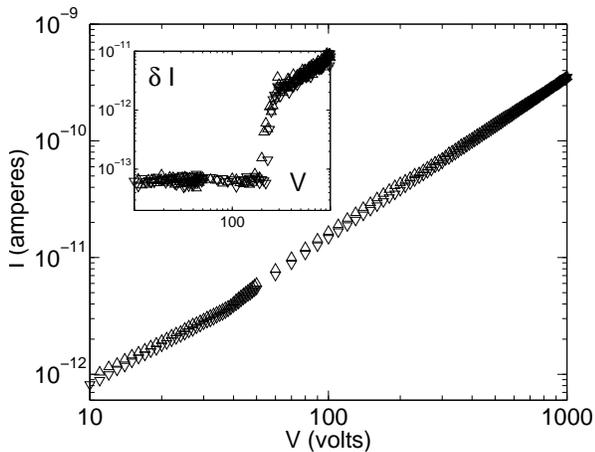} 
\caption{\label{iv}
  A representative current-voltage curve for an annular film
  showing the onset of electroconvection. Data obtained for
  increasing (decreasing) voltages are shown in $\bigtriangleup
  (\bigtriangledown)$. The inset shows the rms fluctuations of the
  current {\it vs.} the applied voltage. Here
 $\Gamma = 3.74 \pm 0.02$ and ${\cal P}=23 \pm 2$. 
 }
\vspace{-0.50cm}
\end{figure}

Fig.~\ref{iv} shows a representative $IV$ response. The critical voltage $V_c$ is identifiable by the upward kink in the $IV$ curve at $V \sim 40$V, after which the film makes a transition from conduction to convection. When $V < V_c$, the fluid is motionless and charge is carried by ohmic conduction. When $V > V_c$, the electrical driving force overcomes dissipation and the film flows in a series of counter-rotating vortices, which are laminar for small $V > V_c$. The fluid circulation carries an additional current by convection, as is apparent by the increase in the slope of the $IV$ curve above $V_c$.  At higher voltages, the vortices become unsteady.  The transition to unsteady flow is identified by a sudden jump in the current fluctuations at $V \sim 200$V, as shown in the inset of Fig.~\ref{iv}. With even higher driving voltage $V \apprge 600$V, the flow becomes turbulent. It is in this highest range that $Nu-{\cal R}$ power law scalings were observed in the experiments.

\subsection{Dependence on the Rayleigh number ${\cal R}$}

The electric Rayleigh number ${\cal R}$, defined by Eqn.~\ref{Rdef}, is a dimensionless measure of the external electric driving force.  Stated in terms of experimentally measurable parameters, ${\cal R} = \epsilon_0 V^2/\sigma_3 \eta_3 s^2$, where $\epsilon_0$ is the permittivity of free space, $V$ is the imposed voltage, $\sigma_3$ is the bulk conductivity, $\eta_3$ is the bulk viscosity, and $s$ is the film thickness. 
We find the bulk conductivity $\sigma_3$ by measuring the ohmic conductance of the film $C = 2\pi s \sigma_3 / \ln{(1/\alpha)}$, where $\alpha$ is the radius ratio $r_i/r_o$.  $C$ can be directly determined from the $IV$ curve, $C = I/V$, in the conductive regime for $V < V_c$. Previous experiments have determined the bulk viscosity to be $\eta_3=0.19\pm 0.05$ kg/ms at atmospheric pressure~\cite{daya_thesis_99}. 

As previously mentioned, we determine the film thickness $s$ by matching the observed reflection color of the film to a color chart. 

The dimensionless Nusselt number $Nu$ measures the convective contribution to the charge transport. $Nu$ can be directly calculated from $IV$ data; $Nu$ is the total current normalized by the conductive current.  The error in the scaling exponent $\gamma$ in the relation $Nu\sim{\cal R}^\gamma$ stems from uncertainties in the film thickness $s$, the critical voltage $V_c$, and the film conductance $C$.  The main source of error is the drift in the film conductance $C$. The drift results in slightly different slopes for the $IV$ curves in the conductive regime between the increasing voltage and decreasing voltage sweeps in a single experiment. The cause of the conductance drift is not well understood, but is presumably due to electrochemical changes in the liquid crystal material. A large drift in the film conductance can lead to a significant uncertainty in the $Nu-{\cal R}$ scaling. In principle, this drift could be completely compensated for if $C$ were known for every $IV$ measurement.  Unfortunately, the conductance $C$ cannot be independently determined from the data while the film is convecting. We must use the $IV$ data in the conduction regime, which occurs before and after the convection regime during one voltage sweep.

In order to bracket the drift, we have analyzed the data using two methods. From a fit to the ohmic response at the beginning and end of each sweep, we determine two conductances, ${C_{up}}$ and ${C_{down}}$.
In method A, we use ${C_{up}}$ to reduce the data obtained when incrementing the voltage and ${C_{down}}$ to reduce the data obtained when decrementing the voltage.   This method concentrates the conductance error at the maximum voltage, and thus overestimates the effect of drift at the highest voltages.  In the second method, which we refer to below as  method B, we instead assume that the conductance $C$ varies linearly between ${C_{up}}$ and ${C_{down}}$ while the fluid is convecting.  This method can be thought of as a linear approximation to the unknown evolution of $C$ while the film is convecting.  It probably underestimates the true drift.  By comparing our results using methods A and B, we can gauge the overall effect of the drift.  We have analyzed all the data using both methods.  In some runs, especially large drifts made compensating for the drift essentially impossible.  We discarded runs in which the difference in ${C_{up}}$ and ${C_{down}}$ exceeded $30\%$.

\begin{figure}
\includegraphics[width=7.5cm]{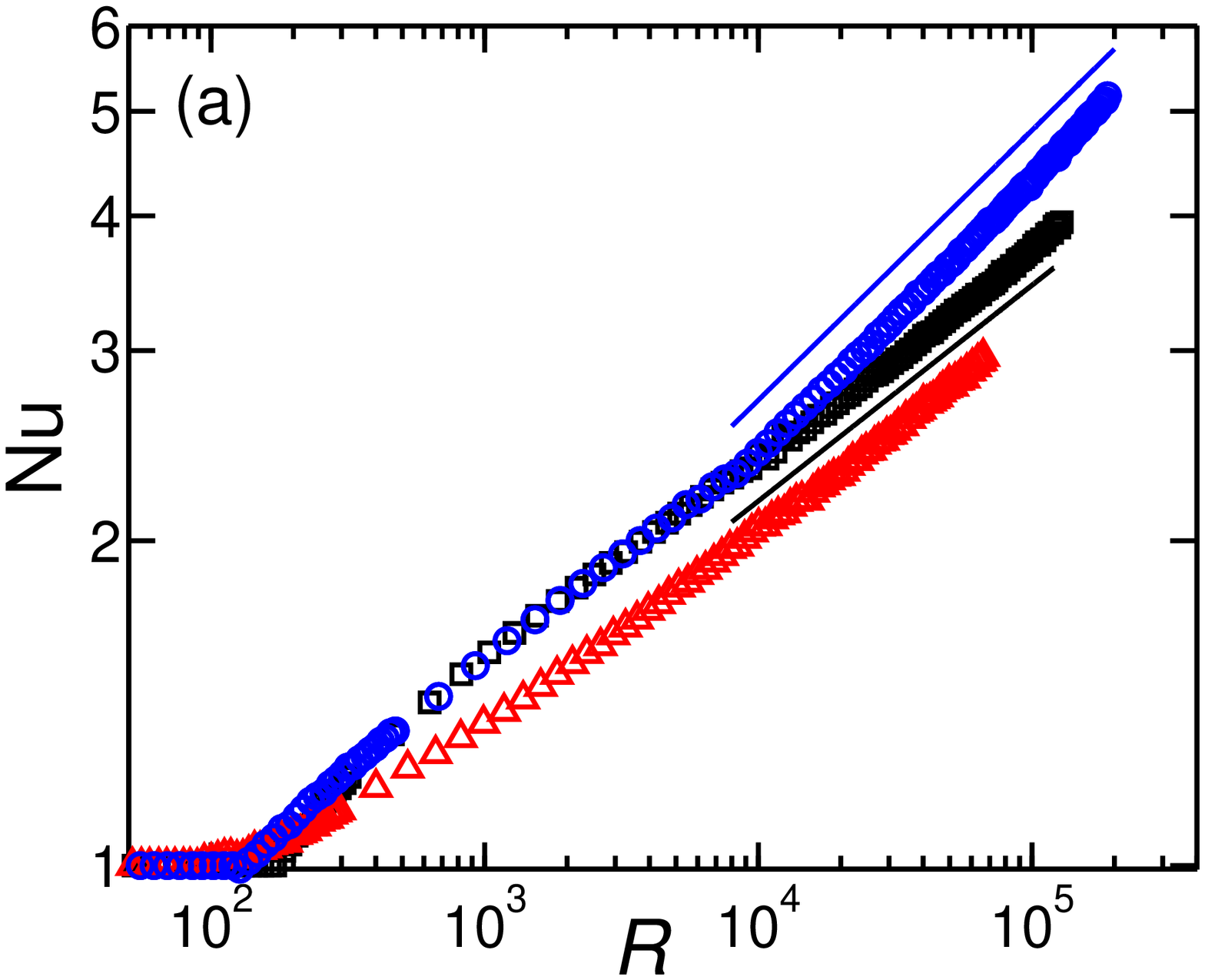}
\includegraphics[width=7.7cm]{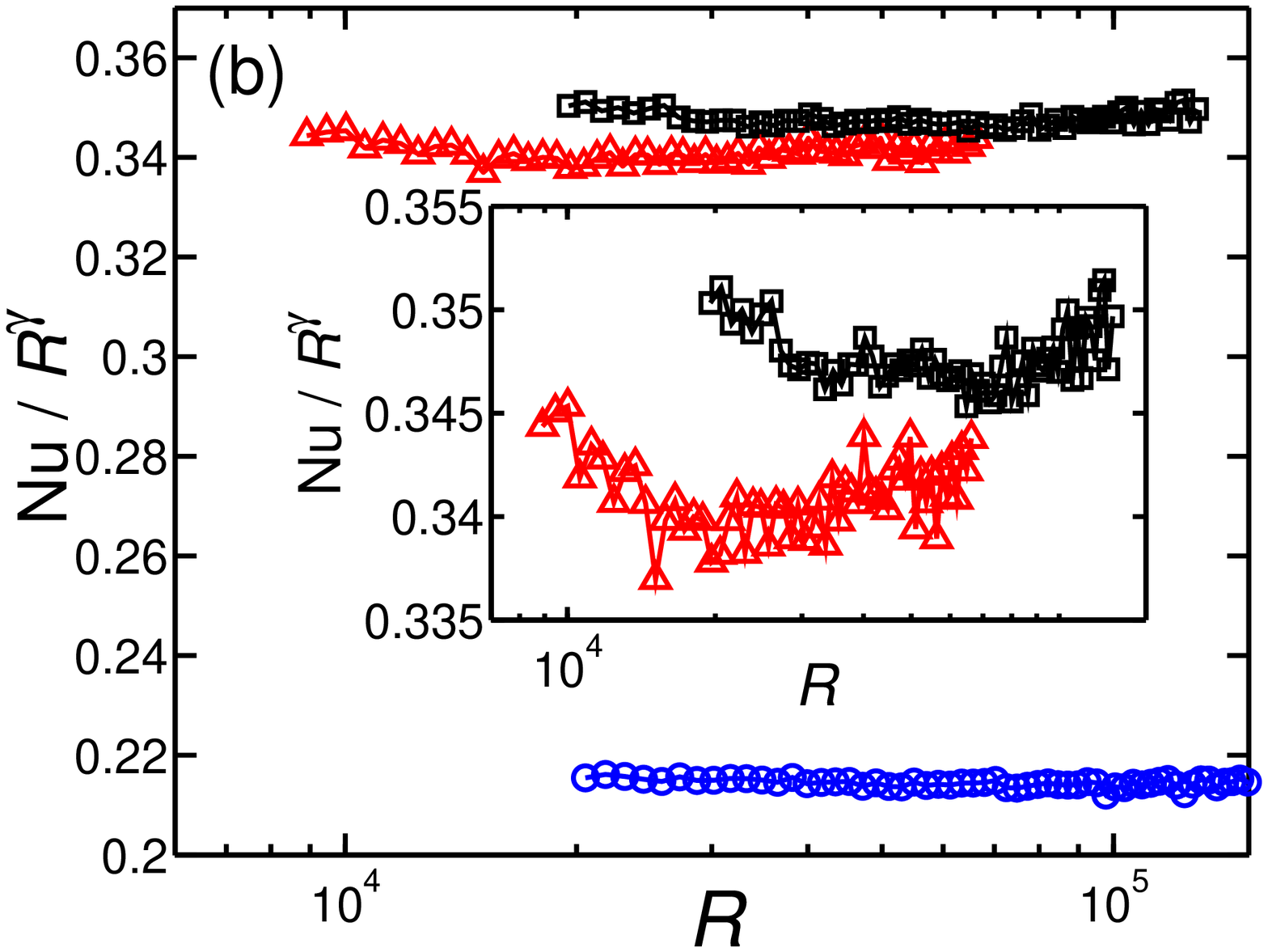}
\caption{\label{NuvsR}
(Color online) (a) shows representative plots of $Nu$ vs ${\cal R}$ for $\Gamma = 0.33~\pm 0.01 ~(\bigtriangleup)$, $3.74~\pm 0.02~(\Box)$, and
$6.60 \pm 0.02~(\circ)$, analyzed by method A. For these, ${\cal P} = 8.8\pm 0.5,~21 \pm 1$, and $~36 \pm 1,$ respectively. 
Least-square fits to the power law $Nu \sim {\cal R}^{\gamma}$ give best fit
values $\gamma =~0.19 \pm 0.01 ,~0.21 \pm 0.02$, and $0.26 \pm 0.01$, respectively. The solid reference lines in the figure have slopes of 1/5 (lower one) and 1/4 (upper one).  In (b) we plot the compensated scaling $Nu~{\cal R}^{-\gamma}$ {\it vs.}  
${\cal R}$ for the same data as in (a). The inset in (b) shows a more expanded scale for $\Gamma = 0.33~(\bigtriangleup)$, and $3.74~(\Box)$.
}
\vspace{-0.20cm}
\end{figure}

Fig.~\ref{NuvsR}a shows $Nu$ {\it vs.} ${\cal R}$ data for three aspect ratios $\Gamma$. When $Nu = 1$, the fluid is quiescent. When $Nu > 1$, the fluid is convecting.  For ${\cal R} \apprge 10^4$, the data reveal power law behavior with $Nu \sim {\cal R}^\gamma$ and values of $\gamma$ close to either $1/5$ (for smaller ${\cal P}$) or $1/4$ (for larger ${\cal P}$), in a good agreement with the theoretical predictions for the small $\cal R$ regime.  Detailed experimental results for various $\Gamma$ and ${\cal P}$ are listed in Table~\ref{table-NuvsR}. The Rayleigh number ${\cal R}$ range used to fit $Nu\sim{\cal R}^\gamma$ was between approximately $100 \times {\cal R}_c$ and the final data point at the largest voltage, $1000$ volts.  Also note that for annular electroconvection, the critical Rayleigh number ${\cal R}_c$ at the onset of convection is $\sim 100$, and thus a factor of $10$ smaller than the corresponding $Ra_c \sim 2000$ for RBC.   From a total of $46$ experiments, the scaling exponents $\gamma$ were either $0.20 \pm 0.03~(0.19 \pm 0.03)$ or $0.25 \pm 0.02~(0.24 \pm 0.02)$ by analysis method A (B), depending on ${\cal P}$. The error bars quoted cover the scatter in $\gamma$ and also include the uncertainties in the film thickness, the critical voltage, and the film conductance drift.  In Fig.~\ref{NuvsR}b  compensated plots of $Nu/{\cal R}^\gamma$ vs. ${\cal R}$ are shown.  The best fit exponent $\gamma$ obtained from experiments was used to compensate the $Nu$ data.  Although only about one decade of scaling range is available, the compensated plots show that a local power law is an adequate description of the data.  This range is probably too short to resolve details about possible cross-over scalings~\cite{GL_01}.

\begin{table}
\caption{\label{table-NuvsR}Results of fits to $Nu \sim {\cal R}^\gamma$, for different aspect ratios $\Gamma$ and various Prandtl numbers ${\cal P}$ by analysis method A. Results obtained by method B are consistent.}
\begin{ruledtabular}
\begin{tabular}{llll}
$\Gamma$ & range of ${\cal P}$ & ~~~~~~$\gamma$ & ~~range of ${\cal R}$\\
\hline
$0.33$ & $6-9$ & $0.21\pm 0.02$ & $8\times 10^3- 7\times 10^4$\\
$1.54$ & $19-28$ & $0.21\pm 0.01$ & $1\times 10^4- 2\times 10^5$\\
$3.74$ & $21-25$ & $0.22\pm 0.02$ & $2\times 10^4- 2\times 10^5$\\
$6.6$ & $25-41$ & $0.24\pm 0.04$ & $2\times 10^4- 2\times 10^5$\\
$6.6$ & $48-61$ & $0.21\pm 0.01$ & $2\times 10^4- 3\times 10^5$\\
$11.1$ & $70-74$ & $0.25\pm 0.02$ & $1 \times 10^4- 2\times 10^5$\\
$11.1$ & $112-120$ & $0.18\pm 0.01$ & $ 2\times 10^4- 4\times 10^5$\\
$11.1$ & $127-136$ & $0.19\pm 0.01$ & $2\times 10^4- 4\times 10^5$\\
$16.1$ & $205-241$ & $0.18\pm 0.02$ & $8\times 10^3- 5\times 10^5$\\
\end{tabular}
\end{ruledtabular}
\end{table}

\subsection{Dependence on the Prandtl number ${\cal P}$}

The electric Prandtl number ${\cal P}$ defined by Eqn.~\ref{Pdef} is the dimensionless ratio of the charge and  viscous relaxation time scales. ${\cal P}^{-1}$ appears as a prefactor in the nonlinear and time derivative terms when the equations of motion are written in dimensionless variables. It is thus reasonable that any dependence on ${\cal P}$ vanishes for large ${\cal P}$. However, for ${\cal P} \sim 1$ the turbulent  flow and scalings are expected to depend on ${\cal P}$. The relative length scales of the electric potential and viscous  boundary layers, which depend on ${\cal P}$, enter into the scaling arguments of Grossmann and Lohse~\cite{GL_00}. For RBC, GL theory predicts that $Nu$ should exhibit local power law scalings with Prandtl number, albeit with rather small powers~\cite{GL_00,GL_01}. RBC experiments suggest that the heat transport is largely independent of the Prandtl number between $4$ to $1350$.  For example, it has been found that $Nu~\sim~{\cal P}^{-0.03}$ at $\Gamma = 1$ ~\cite{xia_lam_zhou_02}. In turbulent electroconvection for $ 2\times 10^4 \leq {\cal R} \leq 10^5$, we find that $Nu$ varies by only a factor of $2$ over the broad range $5 \leq {\cal P} \leq 250$.

\begin{figure}
\begin{center}
\includegraphics[width=7.5cm]{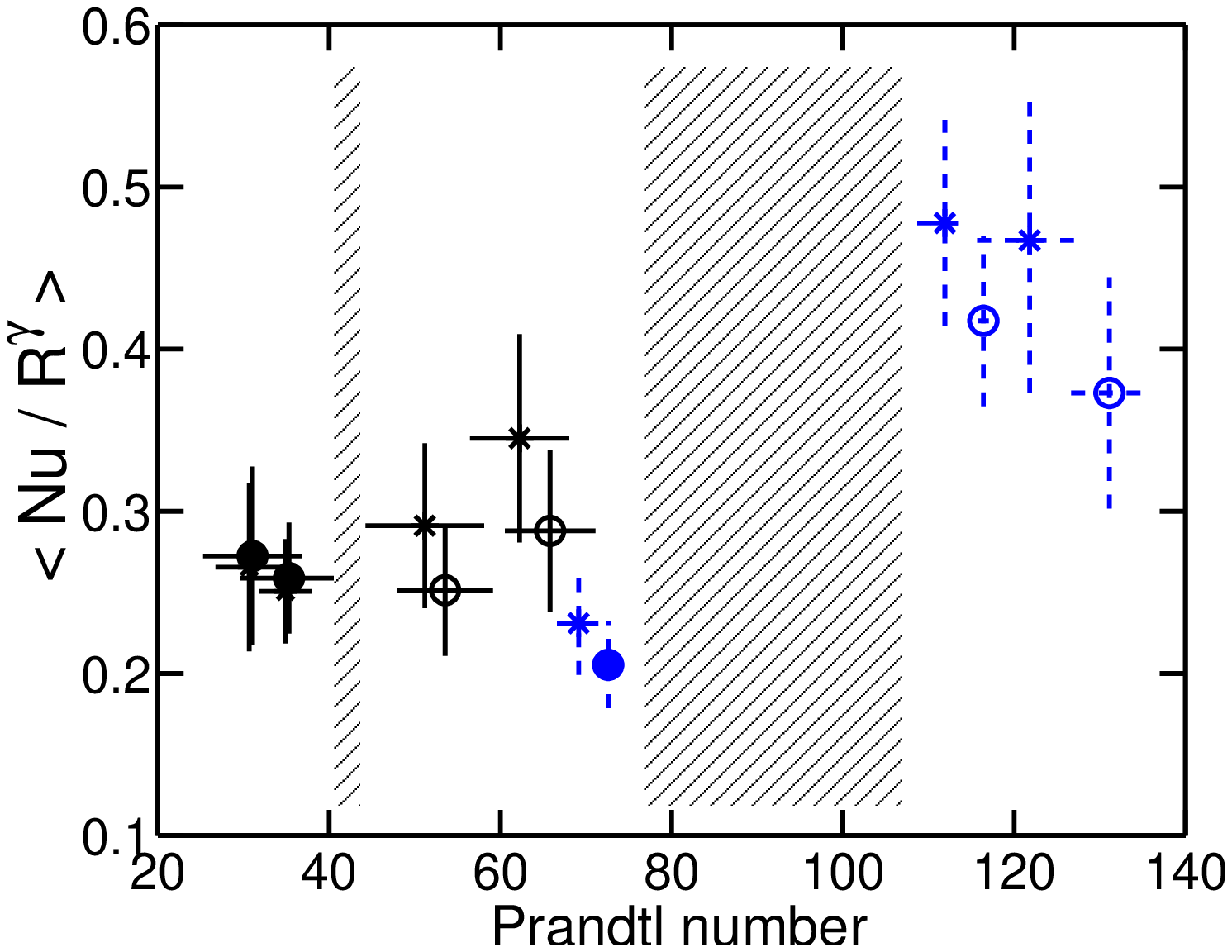}
\end{center}
\caption{\label{NuvsPa}
(Color online) Plots of averaged $Nu/{\cal R}^\gamma$ {\it vs.} ${\cal P}$ for $\Gamma = 6.60 \pm 0.05$~(black symbols) and $11.1 \pm 0.1$~(blue symbols), where $\gamma$ is taken from the best fit to $Nu \sim {\cal R}^\gamma$. Circular symbols $(\circ, \bullet)$ show results obtained by analysis method A, while $(\ast)$ symbols are obtained by method B.  Solid ($\bullet$) and open ($\circ$) symbols indicate when the scaling exponent $\gamma \sim 1/4$~ ($1/5$).   Note that the crossover between $1/4$ and $1/5$  exponents occurs at different ${\cal P}$ for different aspect ratios $\Gamma$, as indicated by the shaded rectangles. 
}
\vspace{0.2cm}
\begin{center}
\includegraphics[width=7.5cm]{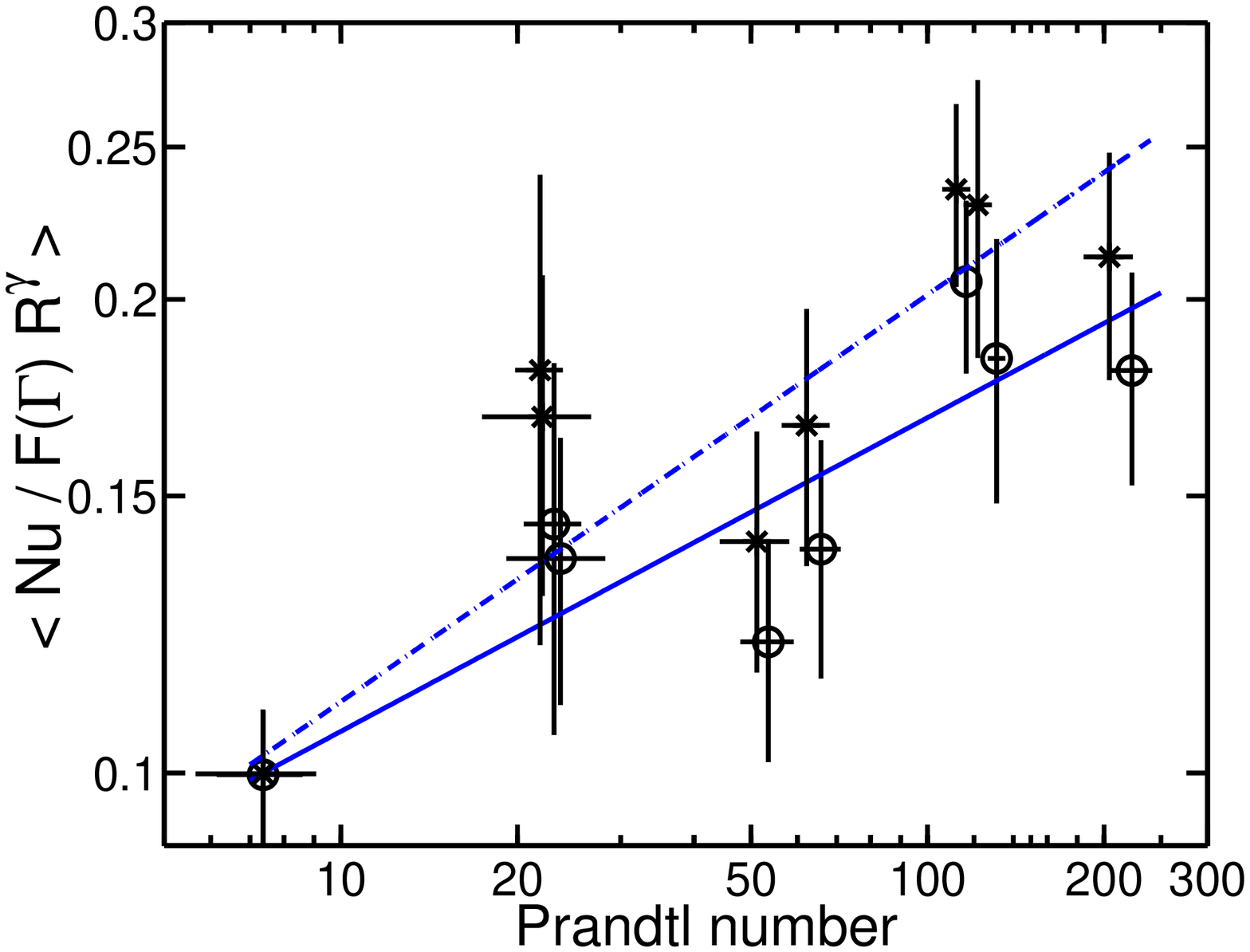}
\end{center}
\caption{\label{NuvsPb}
(Color online) The average value of fully compensated $Nu/({\rm F}(\Gamma) {\cal R}^\gamma$) {\it vs.} ${\cal P}$ on a log-log scale for various $\Gamma$.  Circular symbols $(\circ)$ show results obtained by analysis method A, while $(\ast)$ symbols are obtained by method B.  All of these data had a $Nu$ {\it vs.} ${\cal R}$ scaling exponent of $\gamma \sim 1/5$. The solid (dashed) line is the best fit to $\sim {\cal P}^\beta$ with $\beta = 0.20~(0.26)$, using analysis method A (B). 
}
%\vspace{-0.20cm}
\end{figure}

Fig.~\ref{NuvsPa} shows a plot of $Nu$ {\it vs.} ${\cal P}$ for two aspect ratios $\Gamma = 6.6$ and $11.1$. At each ${\cal P}$, we have averaged the compensated data $Nu{\cal R}^{-\gamma}$ over $2 \times 10^4 \leq {\cal R} \leq 10^5$, with $\gamma$ from the best fit.  The data plotted in Fig.~\ref{NuvsPa} were obtained from a total of $26$ experiments, and results for both analysis methods A and B are shown.

For a fixed aspect ratio $\Gamma$, our data suggest a crossover from one local power law $Nu\sim{\cal R}^\gamma$ to another as ${\cal P}$ increases. Taking $\Gamma = 6.6$ as an example, the $Nu$ {\it vs.} ${\cal R}$ scaling for $20 < {\cal P}< 40$ gives a $\gamma$ exponent  $\approx 1/4$ , while for $50 < {\cal P}< 70$, the exponent is  $ \approx 1/5$. The same indications of a $1/4$ to $1/5$ crossover of scaling exponents are found for $\Gamma = 11.1$, but for a higher value of ${\cal P}$. In the case of $\Gamma = 11.1$, we find $\gamma \approx 1/4$ for $70 < {\cal P} < 80$ and $\gamma \approx 1/5$ for $100 < {\cal P}< 130$. One interpretation of this observation is that the boundaries between different $Nu$ {\it vs.} ${\cal R}$ scaling regimes depend on the aspect ratio $\Gamma$.  This has not been previously considered in scaling theories of RBC, which are specific to the case $\Gamma \sim 1$.  In annular electroconvection, we may have to consider a three dimensional parameter space of $Nu$ scaling regimes that depends on ${\cal R},~{\cal P}$ and $\Gamma$.

The GL scaling theory applied to annular electroconvection yielded an explicit aspect ratio dependence for the dimensionless charge transport $Nu$, given by the function $F(\Gamma)$ defined in Eqn.~\ref{Fdef}. Thus, we can take the varying aspect ratio into account by dividing the $Nu$ data by the theoretically predicted value of $F(\Gamma)$. Fig.\ref{NuvsPb}, shows such a fully compensated plot of $Nu {\cal R}^{-\gamma}{F}^{-1}$ {\it vs.} ${\cal P}$ for various aspect ratios $\Gamma$. Our data span $5 \leq P \leq 250 $ and $0.3 \leq \Gamma \leq 17$. All of this data had $\gamma \sim 1/5$.  A weak dependence on ${\cal P}$ remains, amid considerable scatter.  A power law fit $\sim {\cal P}^\beta$  gives $\beta = 0.20 \pm 0.04$ (analysis method A) and $0.26 \pm 0.05$ (analysis method B).    The theoretical prediction is $Nu \propto F(\Gamma) {\cal R}^{1/5} {\cal P}^{1/5}$ for the regime of low ${\cal R}$ and $Nu$. Thus, our measured value of $\beta$ is at least consistent with the GL theory applied to annular electroconvection for this specific regime, although the scatter is obviously too large to definitively establish that a power law is present.  For the the adjacent regime where $\gamma = 1/4$, we have less than a decade of ${\cal P}$ range which is insufficient even to look for such consistency.

\subsection{Dependence on the aspect ratio $\Gamma$}

In the scaling theory described above, we explicitly accounted for the aspect ratio dependence, a consideration omitted in the GL theory for turbulent RBC~\cite{GL_00,GL_01} which treats the case $\Gamma=1$ . In our formulation of the GL theory for annular electroconvection, we found that the charge transport is modified by an aspect ratio dependent prefactor $\rm{F}(\Gamma)$ given by Eqn.~\ref{Fdef}. Unlike the previous studies of RBC in Ref.~\cite{shr_sig_90_94}, we find the aspect ratio dependence is not a power law scaling but rather a simple function of the annular geometry. To make a direct comparison to previous turbulent RBC experiments, we consider a new function $k\rm{F}$, which we define to be  $\rm{F}(\Gamma)$ multiplied by a normalization constant $k$, chosen such that $k{\rm F}(\Gamma=1)=1$.  The appropriate value of $k$ is $\pi/((\pi+1)\ln(2\pi+1))=0.382$.  The function $k{\rm F}(\Gamma)$ decreases monotonically with $\Gamma$ with its greatest variation for $\Gamma < 2$, and is within $2\%$ of its limiting value $k{\rm F}(\infty)=0.764$  for $\Gamma > 7$.

\begin{figure}
\includegraphics[height=5.5cm]{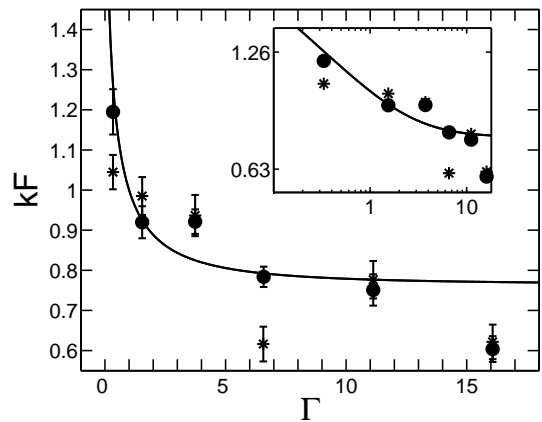}
\caption{\label{Fplot} 
Plot of $k{\rm F}(\Gamma)$ {\it vs.} $\Gamma$ from the scaling theory (solid line), with $k$ chosen so that $k{\rm F}(\Gamma = 1)$ is unity.
Experimental data for turbulent electroconvection are shown by solid symbols ($\bullet$) by method A and by ($\ast$) for method B. The data span $5 \leq {\cal P} \leq 250$. The inset shows the same data on a logarithmic scale.  
}
\vspace{-0.20cm}
\end{figure}

Our experimental data span the range $0.3 < \Gamma < 17$. Because our data span the wide range $5< {\cal P} < 250$, we expect some corrections due to changes in the $Nu$ {\it vs.} ${\cal P}$ scaling, predicted by the theory and also observed in the current experiments. To separate the aspect ratio dependence of $Nu$ and to compare it with the theoretical prediction $Nu \propto {\rm F}(\Gamma){\cal R}^\gamma {\cal P}^\delta$, we divide $Nu$ by ${\cal R}^\gamma{\cal P}^\delta$ and take its averaged value over the ${\cal R}$ range from $2\times 10^4$ to $10^5$.  $Nu$ could still depend on $\Gamma$ independent prefactors which are not captured by a scaling theory.  These prefactors can not be separately extracted from the experimental data and may change for the various scaling regimes~\cite{GL_01}.  To avoid these, we restrict our discussion to those data for which the $Nu$ {\it vs.} ${\cal R}$ scaling exponent $\gamma$ is close to $1/5$. We extract the exponents $\gamma$ from the power law fits of $Nu\sim{\cal R}^\gamma$ to the experimental data. We use $\delta = 0.20$ and $0.26$ which are obtained by the power law fit of $Nu/({\rm F}(\Gamma) {\cal R}^\gamma) \sim {\cal P}^\delta$ shown in Fig.~\ref{NuvsPb}, using the two analysis methods A and B, as described above. This effectively completes a circle of mutual consistency checks which delivers a self-consistent experimental result for the ${\rm F}(\Gamma)$ dependence alone.  Using one free parameter for all the data, we scale these data so that $\rm{F}(\Gamma=1)=1$, to allow a comparison to similarly normalized RBC data.  Our data for six different $\Gamma$, obtained from a total of $36$ experiments, are in a reasonable agreement with the theoretical prediction for $k\rm{F}$, as shown in Fig.~\ref{Fplot}. The errors are representative of the scatter in $Nu/{\cal R}^\gamma {\cal P}^\delta$ among runs for each aspect ratio.

 The data at $\Gamma = 6.6$ and $16.1$, which deviate most from the theoretical prediction for $k{\rm F}$, consist of a few runs which have rather high values of $\cal P$, about $60$ and $220$, respectively.  This deviation may be due to difficulty of finding the appropriate ${\cal P}$ scaling exponent $\delta$.    For relatively small ${\cal R}$ and high ${\cal P}$, scaling theory~\cite{GL_01} predicts that $Nu$ becomes independent of  ${\cal P}$, scaling with ${\cal R}^{1/5} {\cal P}^0$. It is possible that our few data points with very high ${\cal P}$ might fall into the scaling regime $\gamma=1/5,~~\delta=0$ instead of $\gamma=1/5,~~\delta=1/5$, as we have assumed.  Thus, the ${\rm F}(\Gamma)$ data at $\Gamma = 6.6$ and $16.1$ may be significantly underestimated in Fig.~\ref{Fplot}.   Many more experiments with a wider range of parameters would be required to systematically explore the various scaling regimes, boundaries and crossover effects.

Data from several turbulent RBC experiments~\cite{ahlers_00,ahlers_side_00,niemela_00,niemela_03,HKgroup_96} for various values of $\Gamma$ are also in broad agreement with the function $k{\rm F}(\Gamma)$, in spite of the difference in geometry and the higher range of Rayleigh numbers.  Details of this comparison may be found in Ref.~\cite{tsai_04}.  One should only expect the comparison between the aspect ratio dependence for RBC and the function k${\rm F}$ to be reasonable in the limit $\Gamma \rightarrow \infty$. Nevertheless, we find approximate agreement, in spite of the difference in geometry, the smaller aspect ratios and the higher range of Rayleigh numbers.

\section{Conclusion}

We have theoretically and experimentally studied how the dimensionless charge transport $Nu$ scales with the Rayleigh number~${\cal R}$, the Prandtl number~${\cal P}$, and the aspect ratio~${\Gamma}$, for turbulent electroconvection in a 2D annular film. The electroconvection is driven by an unstable charge distribution that is analogous to the inverted fluid density distribution in 3D, thermally driven RBC. The strong similarity of the governing equations between electroconvection and RBC allowed us to adapt GL scaling theory~\cite{GL_00,GL_01} and experimentally investigate its consequences. The unique annular geometry of the electroconvection also made it possible to explicitly account for the aspect ratio dependence of the scaling relations.

From the theory, we found various regime-dependent local power laws of the form $Nu\sim {\rm F}(\Gamma){\cal R}^\gamma {\cal P}^\delta$, with the same exponents as those for turbulent RBC, but with an additional aspect ratio dependent prefactor ${\rm F}(\Gamma)$. ${\rm F}(\Gamma)$ is a nontrivial function of the finite annular geometry, rather than a power law.

In experiments, we found that the exponents for the $Nu$ {\it vs.} ${\cal R}$ scaling were consistent with $1/4$ or $1/5$, for $10^4 \lesssim {\cal R} \lesssim 2\times 10^5$, in reasonable agreement with the theory for the regime where ${\cal R}$ and $Nu$ are both small. The $Nu$ measurements for $0.3 < \Gamma < 17$ and $5 < {\cal P} < 250$ are consistent with the theoretical prediction for the regime where $Nu \sim F(\Gamma) {\cal R}^{1/5}{\cal P}^{1/5}$. Furthermore, our experimental data suggests $\Gamma$-dependent cross-over between different scaling regimes.

 The weak dependence of ${\rm F}$ on $\Gamma$ for large $\Gamma$ suggests that the global heat or charge transport approaches a universal, $\Gamma$ independent limit for laterally extended systems. This conjecture suggests that future work on the charge or heat transport in turbulent convection should focus on large $\Gamma$ systems.  Although such systems can be difficult to achieve experimentally, they are common in nature.

The parameter space of  ${\cal R}$, ${\cal P}$ and $\Gamma$ is, however, very large. In this study, we have have only sparingly sampled from the experimentally accessible portion of this parameter space. In spite of the general consistency between the experiment and theory, it is difficult to draw definitive conclusions about the scaling assumptions underlying the theory. The range of parameters we can access experimentally is too narrow to span the various regimes.  Also, the decomposition of the dissipations into bulk and boundary contributions is not directly testable with our current experimental techniques.

The great strength of the GL scaling analysis is its generality.  It can be applied as easily to our 2D electroconvection as 3D RBC.  The relative simplicity of the 2D fluid mechanics in electroconvection suggest that we can greatly extend the range of parameters by numerical simulation.  Simulations will also allow us to test directly the scaling assumptions.  The GL theory could also be extended to the case of convection with a superposed shear~\cite{daya_98,daya_99,langer_01,daya_01,daya_02}, a situation which is experimentally feasible in electroconvection but not in RBC. 

The nearly power law scaling of the globally averaged heat or charge transport in turbulent convection poses an interesting and difficult problem.    Any approach to this problem must creatively  combine theory, simulation and precision experiment.  The unique features of thin film electroconvection give us a new vantage point on this challenging, and as yet unsolved problem.

\acknowledgements
This research was supported by the Natural Science and Engineering Research Council of Canada.

\end{document}